\documentclass[aip,amsmath,amssymv,reprint]{revtex4-2}
\usepackage{graphicx}
\usepackage{float}
\usepackage{subfig}
\graphicspath{{./}}

\usepackage{dcolumn}
\usepackage{bm}

\usepackage[utf8]{inputenc}
\usepackage[T1]{fontenc}
\usepackage{mathptmx}
\usepackage{etoolbox}

\usepackage[version=4]{mhchem}
\usepackage{braket}
\usepackage{multirow} 
\usepackage{tabularray}

\usepackage{subfiles}
\usepackage{todonotes}
\usepackage{amsmath, amssymb}
\usepackage{ulem}
\usepackage{orcidlink}
\usepackage{lipsum}

\newcommand{\Endo}[2]{\ce{#1}@\ce{#2}}
\newcommand{\wavenumber}{cm$^{-1}$}

\makeatletter
\def\@email#1#2{%
 \endgroup
 \patchcmd{\titleblock@produce}
  {\frontmatter@RRAPformat}
  {\frontmatter@RRAPformat{\produce@RRAP{*#1\href{mailto:#2}{#2}}}\frontmatter@RRAPformat}
  {}{}
}%
\makeatother

\begin{document}
    \title[\Endo{Ne}{C70}]{Application of Correlated-Wavefunction and Density-Functional Theories to Endofullerenes: A Cautionary Tale} 

    \author{K. Panchagnula*\orcidlink{0009-0004-3952-073X}}
    \email{ksp31@cantab.ac.uk}
        \affiliation{Yusuf Hamied Department of Chemistry, University of Cambridge, Cambridge, United Kingdom}
    \author{D. Graf \orcidlink{0000-0002-7640-4162}}
        \affiliation{Department of Chemistry, University of Munich (LMU), Munich, Germany}
    \author{K.R. Bryenton \orcidlink{0000-0001-5716-5314}}
        \affiliation{Department of Physics and Atmospheric Science, Dalhousie University, 6310 Coburg Road, Halifax, Nova Scotia, Canada}
    \author{D.P. Tew \orcidlink{0000-0002-3220-4177}}
        \affiliation{Physical and Theoretical Chemistry Laboratory, University of Oxford, South Parks Road, United Kingdom}
    \author{E.R. Johnson \orcidlink{0000-0002-5651-468X}}
        \affiliation{Yusuf Hamied Department of Chemistry, University of Cambridge, Cambridge, United Kingdom}
        \affiliation{Department of Physics and Atmospheric Science, Dalhousie University, 6310 Coburg Road, Halifax, Nova Scotia, Canada}
        \affiliation{Department of Chemistry, Dalhousie University, 6243 Alumni Crescent, Halifax, Nova Scotia, Canada}
    \author{A.J.W. Thom \orcidlink{0000-0002-2417-7869}}
        \affiliation{Yusuf Hamied Department of Chemistry, University of Cambridge, Cambridge, United Kingdom}

    \date{\today}

    \begin{abstract}
        A recent study by Panchagnula \textit{et al.} [J. Chem. Phys. \textbf{161}, 054308 (2024)] illustrated the non-concordance of a variety of electronic structure methods at describing the symmetric double-well potential expected along the anisotropic direction of the endofullerene \Endo{Ne}{C70}. In this article we present new correlated-wavefunction data from coupled-cluster theory for this system, and scrutinise a variety of state-of-the-art density-functional approximations (DFAs) and dispersion corrections (DCs). We identify rigorous criteria for the double-well potential and compare the shapes, barrier heights, and minima positions obtained with the DFAs and DCs to the correlated wavefunction data. We show that many of the DFAs are extremely sensitive to the numerical integration grid used, the dispersion damping function, and the extent of exact-exchange mixing. 
        We pose the \Endo{Ne}{C70} system as a challenge to functional developers and as a diagnostic system for testing dispersion corrections, and reiterate the need for more experimental data for comparison.

    \end{abstract}

    \maketitle

    \section{Introduction \label{sec: Intro}}

        Non-covalent interactions, particularly London dispersion forces, pose a well-known challenge for \textit{ab initio} electronic structure (ES) methods\cite{grimmeDispersionCorrectedMeanFieldElectronic2016a} due to their weak and long-ranged nature. Therefore, chemical species primarily stabilized by dispersion interactions are often difficult to describe accurately with these methods. 

       Due to its favourable cost-to-performance ratio, density-functional theory (DFT) has earned itself the title of being the ``workhorse of quantum chemistry''\cite{tealeDFTExchangeSharing2022} and has had renowned success across various disciplines. However, its quality is tied to the choice of density-functional approximation (DFA),\cite{perdewJacobsLadderDensity2001} almost all of which have severe limitations in their ability to accurately capture dispersion interactions. In order to alleviate this, a variety of dispersion corrections (DCs) have been developed,\cite{klimesPerspectiveAdvancesChallenges2012} with varying levels of empiricism. These include Grimme's dispersion corrections,\cite{grimmeAccurateDescriptionVan2004, grimmeConsistentAccurateInitio2010, grimmeImprovedSecondorderMoller2003, grimmeSemiempiricalGGAtypeDensity2006, caldeweyherExtensionD3Dispersion2017, caldeweyherGenerallyApplicableAtomiccharge2019} as well as the VV10,\cite{vydrovNonlocalVanWaals2010, sabatiniNonlocalVanWaals2013} many-body dispersion (MBD),\cite{ambrosettiLongrangeCorrelationEnergy2014, hermannDensityFunctionalModel2020} and exchange-hole dipole moment (XDM) \cite{johnsonPostHartreeFockModelIntermolecular2005, beckeExchangeholeDipoleMoment2007, priceXDMcorrectedHybridDFT2023} methods. Grimme's models, in particular, have gained enormous popularity due to their simple and efficient implementation. 

       Correlated wavefunction (WF) methods are an alternative to DFAs that are reputed for their high accuracy, albeit at the expense of a significant increase in computational cost due to unfavourable scaling with respect to both system and basis-set sizes. As dispersion forces explicitly arise due to the interactions of multiple electrons, WF methods account for this in their description of electron correlation. Second order M\o{}ller--Plesset perturbation theory (MP2) is a standout choice due to its efficient implementations,\cite{pulayLocalizabilityDynamicElectron1983, haserMollerPlessetMP2Perturbation1993, maslenNoniterativeLocalSecond1998,ayalaLinearScalingSecondorder1999,schutzLoworderScalingLocal1999, saeboLowscalingMethodSecond2001, wernerFastLinearScaling2003, jungFastCorrelatedElectronic2006,jungFastEvaluationScaled2007, doserTighterMultipolebasedIntegral2008, doserLinearscalingAtomicOrbitalbased2009, zienauCholeskydecomposedDensitiesLaplacebased2009, kristensenMP2EnergyDensity2012, maurerCholeskydecomposedDensityMP22014, pinskiSparseMapsSystematic2015,nagyIntegralDirectLinearScalingSecondOrder2016, baudinEfficientLinearscalingSecondorder2016, phamHybridDistributedShared2019, barcaQMP2OSMollerPlesset2020, glasbrennerEfficientReducedScalingSecondOrder2020, forsterQuadraticPairAtomic2020} whereas coupled cluster (CC) including single, double, and perturbative triple excitations (CCSD(T)) is nominally considered the ``gold standard'' of WF methods.\cite{nagyStateoftheartLocalCorrelation2024} This impressive level of accuracy comes at a heavy price, as CCSD(T) calculations with moderate basis sets tend to be computationally unfeasible for all but very small chemical systems. 

        Between the multiple pairings of DFAs and DCs, as well as  WF methods, there is a plethora of ES methods that could be used to describe the dispersion interaction within a chemical system. Given an accuracy tolerance, it is not immediately obvious as to which choice can achieve it, and for the lowest computational cost. In order to test, validate, and benchmark ES methods for their ability to describe dispersion interactions accurately and efficiently, endofullerenes (EFs) emerge as interesting candidates.

        EFs are a class of systems where atom(s) or molecule(s), \ce{A}, are trapped within a fullerene cage \ce{C$_n$}, denoted \Endo{A}{C$_n$}.\cite{bacicPerspectiveAccurateTreatment2018} The development of a technique known as ``molecular surgery'' has allowed for the controlled synthesis and characterisation of these species, leading to a vast amount of very precise spectroscopic data.\cite{bloodworthSynthesisEndohedralFullerenes2022, levittSpectroscopyLightmoleculeEndofullerenes2013} The sizes of the fullerene rings are such that London dispersion is a significant component of their interaction energies with the encapsulated species. Moreover, they are not too small such that the ES calculations are trivial, and not too large that large basis sets are unfeasible.\cite{cioslowskiElectronicStructureCalculations2023} 

        Spectroscopic data for these systems probes information about the nuclear energy levels (translational, rotational etc.) of the endohedral species measured at the wavenumber level. In order to achieve this level of accuracy theoretically, the potential energy surface (PES) derived from ES calculations must also be at least as accurate. However, the development of high level \textit{ab intio} ES techniques is usually focused on the goal of achieving ``chemical accuracy'' of 1~kcal/mol, which is approximately 350~\wavenumber. Evidently, requiring the ES methods to move from thermochemical accuracy to spectroscopic accuracy, at the 1~\wavenumber\, level, significantly increases the computational demand.

        A recent theoretical study of \Endo{He}{C60}\cite{panchagnulaTranslationalEigenstatesHe2024a} found that MP2 achieved this level of accuracy, with the random phase approximation (RPA) following closely behind. Experimental data for \Endo{He}{C60}\cite{bacanuExperimentalDeterminationInteraction2021a, jafariTerahertzSpectroscopyHelium2022} 
        was available for comparison, 
        which guided the categorisation of the ES methods, but equivalent data was not present for \Endo{Ne}{C70}.\cite{panchagnulaTargetingSpectroscopicAccuracy2024} This larger system proved much more challenging as, even greatly expanding the number and quality of the ES methods considered, a worrying lack of concordance between them was noted. The crux of the difference seemed to lie in the description of the symmetric double well potential along the unique, anisotropic direction of \ce{C70}, with MP2 being a standalone outlier. 
        
        

        In this paper, we delve deeper into the cornucopia of possible ES methods to gain an intuition and understanding of their behaviour for \Endo{Ne}{C70}. As in our previous work, we investigate a one-dimensional slice of the \Endo{Ne}{C70} PES along the unique, anisotropic axis of \ce{C70}, and assess its double-well characteristics. We primarily examine the behaviour of various DFAs, as well as different DCs, to understand what properties are important for an accurate description of the endohedral interaction. We compare the characteristics of these PESs to the previously generated correlated WF data, alongside a few new state-of-the-art CC calculations.

    \section{Theory \label{sec: Methodology}}

        Previous research on PESs for EFs has tended to forego ES calculations,  instead approximating the surface using a pairwise additive Lennard-Jones (LJ) potential, summed over cage-endohedral sites.\cite{xuH2HDD22008, xuQuantumDynamicsCoupled2008,xuCoupledTranslationrotationEigenstates2009, xuInelasticNeutronScattering2013, felkerCommunicationQuantumSixdimensional2016, xuLightMoleculesNanocavities2020, felkerFlexibleWaterMolecule2020, jafariNeArKr2023, panchagnulaExploringParameterSpace2023a, mandziukQuantumThreeDimensional1994} This has the advantage of computational simplicity and ensures that the PES has the appropriate smoothness and symmetry, but at the expense of using empirically derived LJ parameters that may not be optimal.\cite{xuCoupledTranslationrotationEigenstates2009, panchagnulaExploringParameterSpace2023a}
        On the other hand, due to the high computational cost, generating the full PES from ES calculations requires interpolation between a small set of discrete data points.\cite{panchagnulaTargetingSpectroscopicAccuracy2024} Therefore, we require the PES obtained from an ES method to be smooth and differentiable. 
       
       In the specific case of \Endo{Ne}{C70}, we expect to see a well-behaved symmetric double well, without any oscillations. More rigorously, we require the origin to be a local maximum, with only two inflection points present. This criterion ensures there are no extra oscillations, nor any shoulders nor plateaus, which could cause failures in the subsequent calculations of vibrational frequencies, intensities, and other derived properties.\cite{sitkiewiczHowReliableAre2022, sitkiewiczSpuriousOscillationsCaused2024} These requirements pose a particular challenge for DFAs, as the energy is usually evaluated on integration grids prone to numerical inaccuracies.\cite{johnsonApplication25Density2004, johnsonOscillationsMetageneralizedgradientApproximation2009, jimenez-hoyosEvaluationRangeseparatedHybrid2008, csonkaEvaluationDensityFunctionals2009, wheelerIntegrationGridErrors2010, dasguptaStandardGridsHighprecision2017, morganteDevilDetailsTutorial2020, mardirossianHowAccurateAre2016, gouldAreDispersionCorrections2018} In the following, we will give a brief summary of the most important aspects of DFT and the associated numerical integration of the energy functionals, while the specific computational details are provided in the supplementary information.

        In (Kohn--Sham)-DFT, a system's total energy is given by\cite{hohenbergInhomogeneousElectronGas1964, kohnSelfConsistentEquationsIncluding1965a}
        \begin{equation}
            E_\text{DFT} = T_0 + \int v_{\text{ext}} (\bm{r}) \rho(\bm{r}) \, d\bm{r} + E_{\text{H}}[\rho] + E_{\text{XC}} \,,
        \end{equation}
        where $T_0$ is the non-interacting kinetic energy of the electrons, $v_{\text{ext}}$ is the external potential, $E_{\text{H}}$ is the Hartree energy, and $E_{\text{XC}}$ is the exchange-correlation (XC) energy. The XC-energy includes all remaining contributions to the total energy and is hence at the heart of all common DFAs. A large array of different DFAs exists in the literature, with Perdew's ladder \cite{perdewJacobsLadderDensity2001} providing a classification scheme that groups DFAs into five different ``rungs'' according to the ingredients used in approximating the XC-energy. 
        
        In this work, we employed various DFAs from four of the five rungs and compared their performance on the chosen slice of the \Endo{Ne}{C70} PES. From rung 2, we employed the generalized gradient approximation (GGA) functional of Perdew--Burke--Ernzerhof (PBE)\cite{perdewGeneralizedGradientApproximation1996, perdewGeneralizedGradientApproximation1997} and B86bPBE,\cite{beckeLargegradientBehaviorDensity1986} which combines Becke's B86b exchange functional with PBE correlation. From rung 3, we picked the combinatorially-optimized meta-GGA (mGGA) including VV10 dispersion from the Head-Gordon group, called B97M-V.\cite{mardirossianMappingGenomeMetageneralized2015} From rung 4, we employed hybrid variants of PBE and B86bPBE, termed PBE0 \cite{adamoReliableDensityFunctional1999} and B86bPBE0,\cite{priceXDMcorrectedHybridDFT2023} with and without various dispersion corrections (\textit{vide infra}). We additionally considered Becke's 10-parameter global hybrid, B97,\cite{beckeDensityfunctionalThermochemistrySystematic1997} and the following combinatorially-optimized range-separated hybrids from the Head-Gordon group: $\omega$-B97,\cite{chaiSystematicOptimizationLongrange2008} $\omega$-B97X,\cite{chaiSystematicOptimizationLongrange2008} $\omega$-B97X-V,\cite{mardirossianOB97XV10parameterRangeseparated2014} and $\omega$-B97M-V.\cite{mardirossianOB97MVCombinatoriallyOptimized2016} Finally, from rung 5, we chose the double-hybrid B2PLYP\cite{grimmeSemiempiricalHybridDensity2006} and PWPB95\cite{goerigkEfficientAccurateDoubleHybridMetaGGA2011} functionals from the Grimme group, as well the $\omega$-B97M(2)\cite{mardirossianSurvivalMostTransferable2018} functional from the Head-Gordon group. The functionals used were mainly chosen because of their high popularity and excellent performance in typical benchmarks.\cite{goerigkLookDensityFunctional2017}

       Many XC functionals neglect the energy contribution from dispersion effects, so a correction term must be added. Usually, the contribution of dispersion to the electron density is negligible; therefore, a popular approach is to include dispersion as a post-self-consistent-field (post-SCF) correction to the total DFT energy as
        \begin{equation}
            E_{\text{Total}} = E_{\text{DFT}} + E_{\text{disp}} \,.\label{eq: DFT+disp}
        \end{equation}
        Various post-SCF dispersion correction methods have been developed that depend on different system properties. For instance, the Grimme DFT-D series, such as D3(0) \cite{grimmeConsistentAccurateInitio2010} and D3(BJ), \cite{grimmeEffectDampingFunction2011} depend on coordination number. The term in parentheses indicates the specific damping function applied to the DC, ensuring appropriate asymptotic behaviour by making the dispersion energy become constant at small internuclear separations. These functions contain tunable parameters that are optimised for each functional and sometimes basis set. D3(0) and D3(BJ) use the Chai and Head-Gordon \cite{chaiLongrangeCorrectedHybrid2008} and Becke--Johnson (BJ)\cite{johnsonPostHartreeFockModelIntermolecular2006a} damping functions, respectively. The Tkatchenko--Scheffler (TS) \cite{tkatchenkoAccurateMolecularVan2009} and many-body dispersion (MBD)-based models, such as the range-separated self-consistent screening (MBD@rsSCS) \cite{tkatchenkoAccurateEfficientMethod2012, ambrosettiLongrangeCorrelationEnergy2014} and non-local (MBD-NL) \cite{hermannDensityFunctionalModel2020} variants, depend on the electron density and use Fermi-type Wu--Yang damping functions.\cite{wuEmpiricalCorrectionDensity2002} The XDM model \cite{beckeExchangeholeDipoleMoment2007, johnsonChapter5ExchangeHole2017, priceXDMcorrectedHybridDFT2023} depends not only on the density, but also its gradient, Laplacian, and the kinetic-energy density, and uses either the BJ damping function\cite{johnsonPostHartreeFockModelIntermolecular2006a} or a newly proposed Z damping function.\cite{dh24,xcdmz}
        
        In this work, we considered the PBE0 hybrid functional paired with each of the D3(0), D3(BJ), TS, MBD@rsSCS, MBD-NL, XDM(BJ), and XDM(Z) dispersion methods. The two XDM variants were also paired with the B86bPBE0 and B86bPBE50 hybrids (as B86b is better able to describe non-bonded repulsion due to the adherence of its enhancement factor to the appropriate large reduced-density-gradient limit\cite{priceRequirementsAccurateDispersioncorrected2021}), as well as the LC-$\omega$PBEh range-separated hybrid.
        
        As the mixing parameters for the MP2 contribution in the B2PLYP and PWPB95 double-hybrid functionals are smaller than unity, they were combined with D3(BJ) to ensure a asymptotically correct description of long-range dispersion effects.\cite{goerigkEfficientAccurateDoubleHybridMetaGGA2011, schwabeDoublehybridDensityFunctionals2007}
        Note that the B97M-V, $\omega$-B97X-V, $\omega$-B97M-V, and $\omega$-B97M(2) functionals were not paired with another dispersion correction as they already include the VV10 dispersion method in their definitions. While $\omega$-B97X does not include a dispersion correction, it was parameterized to reference data involving dispersion-bound systems and, therefore, captures some short-range dispersion-like binding through its exchange terms.

        Although the chosen DFAs exhibit significantly different functional forms, they share the common characteristic that their energies are evaluated on a numerical integration grid. Typically, the molecular grid is decomposed into a collection of atom-centered grids using Becke's partitioning \cite{beckeMulticenterNumericalIntegration1988} or other similar schemes.\cite{laquaImprovedMolecularPartitioning2018a,treutlerEfficientMolecularNumerical1995} Each atomic grid is generally constructed as a product of radial and angular grids, denoted as a tuple of (radial points, angular points) used, with the angular points corresponding to the Lebedev quadratures.\cite{lebedevQUADRATUREFORMULASPHERE1999}
        The behaviour of functionals  for intermolecular interactions can be extremely sensitive to the integration grid. Too small a grid can not only mean that certain integrals do not converge, but in the case of certain DFAs, introduce spurious oscillations into the PES, essentially rendering it useless.\cite{johnsonApplication25Density2004,johnsonOscillationsMetageneralizedgradientApproximation2009, jimenez-hoyosEvaluationRangeseparatedHybrid2008, csonkaEvaluationDensityFunctionals2009, wheelerIntegrationGridErrors2010, dasguptaStandardGridsHighprecision2017, morganteDevilDetailsTutorial2020, mardirossianHowAccurateAre2016, sitkiewiczHowReliableAre2022, sitkiewiczSpuriousOscillationsCaused2024} On the other hand, too large a grid can severely increase the computational cost, making the calculation infeasible. 
        The sensitivity of the computed \Endo{Ne}{C70} PES to the integration grid is illustrated in the SI for the case of $\omega$-B97M-V and, without extremely dense radial integration grids, the PES obtained with certain functionals exhibited strong oscillations.

        The domain-based local pair natural orbital (DLPNO)-CCSD and DPLNO-CCSD(T0) methods\cite{neeseEfficientAccurateLocal2009,guoCommunicationImprovedLinear2018,liakosComprehensiveBenchmarkResults2020,franzkeTURBOMOLETodayTomorrow2023} approximate equivalent canonical CCSD and CCSD(T) calculations by expanding correlation in pair natural orbital (PNO) spaces defined from MP2 pair densities. Local domains and pair-specific truncations systematically compress the virtual space, with accuracy controlled by a threshold parameter.\cite{tewPrincipalDomainsLocal2019}  Power-law extrapolation with both basis set size and threshold parameter leads to an energy estimate with error bars from extrapolation.\cite{sorathiaImprovedCPSCBS2024}
    
    \section{Results \label{sec: Results}}

        \begin{figure*}
            \centering
            \subfloat[WF]{\includegraphics[width=0.49\textwidth]{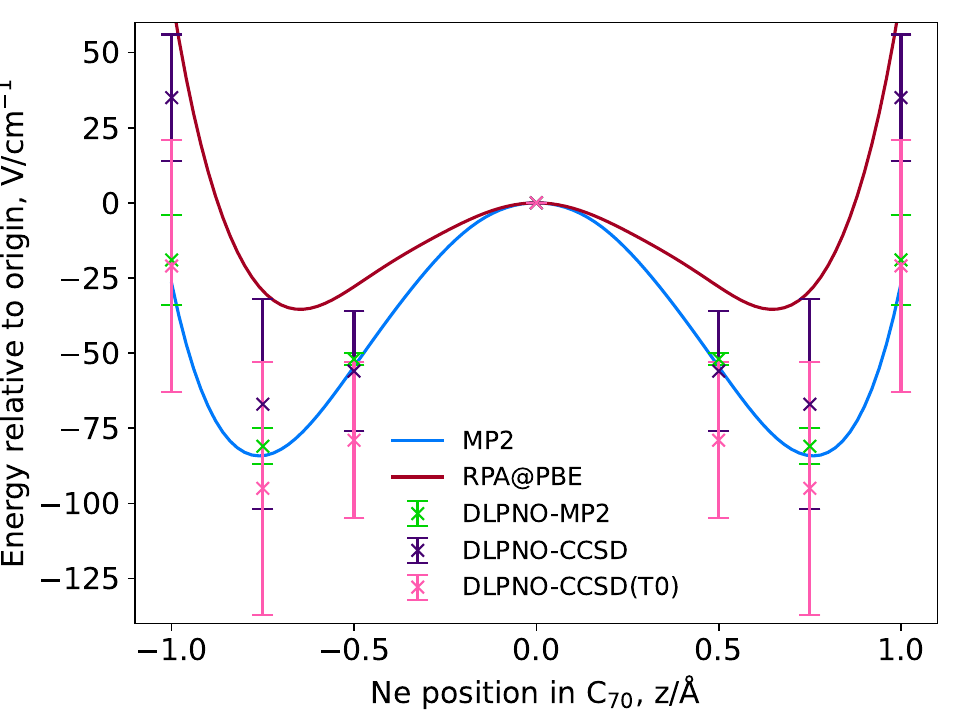} \label{fig: WF}}
            \hfill
            \subfloat[DFA]{\includegraphics[width=0.49\textwidth]{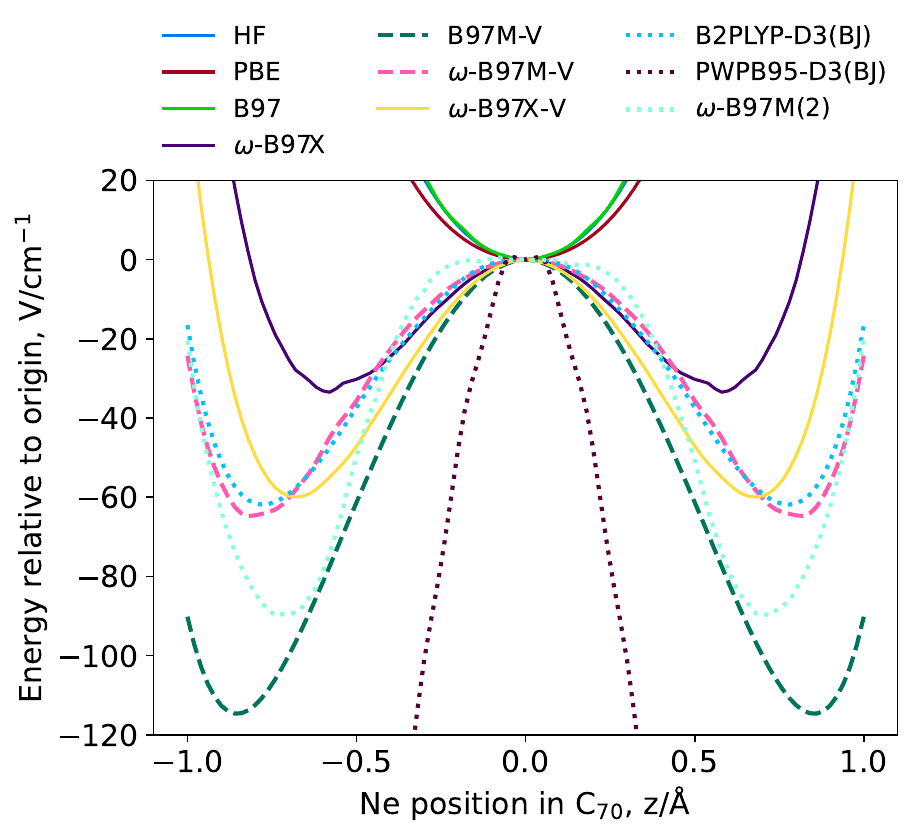} \label{fig: DFA}}\,
            \caption{
            PES slices of (a) WF methods and (b) DFAs for points separated by 0.02\AA. For the WF methods, MP2 is shown in blue and RPA@PBE in maroon; DLPNO single points are shown as crosses with error bars with MP2, CCSD and CCSD(T0) in green, purple and pink respectively.
            For the DFAs, GGAs and GGA-based hybrids are given by solid lines, mGGA-based hybrids by dashed lines, and double hybrids by dotted lines. HF, PBE, B97, $\omega$-B97X, B97M-V, $\omega$-B97M-V, $\omega$-B97X-V, B2PLYP-D3(BJ), PWPB95-D3(BJ), $\omega$-B97M(2) are in blue, maroon, light green, purple, forest green, pink, yellow, light blue, brown, and turquoise.
            }
            \label{fig: Functionals Behaviour}
        \end{figure*}
        \begin{table}
            \caption{Barrier heights and minima positions for different DFAs and WF methods in Figure \ref{fig: Functionals Behaviour}. The DLPNO based WF methods are grouped together as they are single-point calculations with large error bars.}
            \label{table: Functionals Behaviour}
            \centering
            \setlength{\tabcolsep}{8pt}
            \begin{tabular}{ccc}
                 \hline Method&BH/\wavenumber&$z_m$/\AA  \\
                 \hline HF&0&0\\ 
                 PBE&0&0\\ 
                 B97&0&0\\ 
                 $\omega$-B97X   &~~33.48&0.58\\ 
                 B97M-V          &114.64&0.86\\ 
                 $\omega$-B97X-V &~~59.94&0.68\\ 
                 $\omega$-B97M-V &~~64.71&0.82\\ 
                 B2PLYP-D3(BJ)   &~~61.85&0.78\\ 
                 $\omega$-B97M(2)&~~89.84&0.74\\ 
                 PWPB95-D3(BJ)   &293.43&0.74\\ 
                 \hline
                 MP2             &~~84.21&0.76\\ 
                 RPA@PBE         &~~35.43&0.64\\ 
                 DLPNO-WF  & 67--95 & 0.60--0.80 \\
                 \hline
            \end{tabular}
            
        \end{table}
        
        Two metrics that will be used to describe the overall shape of the PES are the minima positions ($z_m$, in \AA) and the  barrier height (BH, in \wavenumber) connecting them; the magnitude of the BH exactly corresponds to the well depth since the zero of energy is defined to be the point at $z=0$ with the \ce{Ne} at the centre of the \ce{C70} cage. The new DLPNO-WF data suggests that the true BH and $z_m$ are likely to lie much closer to those predicted by MP2 than RPA@PBE, and this should serve as a reference when considering the results of the various DFAs. The DPLNO-WF methods are in approximate concordance within the error bars of extrapolation, suggesting there is a barrier, with a likely height on the order 67-95\wavenumber.   The significant time required to perform these calculations puts these results at the limit of present feasibility.

        Figure~\ref{fig: DFA} shows results for the \Endo{Ne}{C70} PES obtained with the selected GGAs and GGA-based hybrids (solid lines), mGGA-based hybrids (dashed lines), and double hybrids (dotted lines).  
        Table~\ref{table: Functionals Behaviour} further summarises the obtained minima positions and barrier heights with selected methods. Dispersion-uncorrected HF, PBE, and B97 expectedly display only a single well  while the remaining methods all display double wells, but of varying characters. $\omega$-B97X gives the shallowest double well of the DFAs considered, likely as it also does not include an explicit dispersion correction. Conversely, the B97M-V and PWPB95-D3(BJ) curves exhibit even deeper double wells than obtained with MP2, with the PWPB95-D3(BJ) double hybrid being a severe outlier with a barrier of nearly 300~\wavenumber. This latter PES also shows some very subtle wiggles, which is likely caused by the high sensitivity of the underlying B95 mGGA to grid size.\cite{johnsonPostHartreeFockModelIntermolecular2005, johnsonOscillationsMetageneralizedgradientApproximation2009} 
        The functionals providing PES with the most comparable features to the WF results are $\omega$-B97M-V, B2PLYP-D3(BJ), and $\omega$-B97X-V. While $\omega$-B97M(2) gives a BH in the closest agreement with the DLPNO-WF references, the corresponding PES shows a flat shoulder (with some very shallow oscillations) near $z=0$ that is not present with any of the other methods. Thus, functionals that contain many parameters, and are traditionally designed for thermochemistry, may not necessarily perform well for this PES.

        \begin{figure*}
            \centering
            \subfloat[PBE0+Dispersion]{\includegraphics[width=0.49\textwidth]{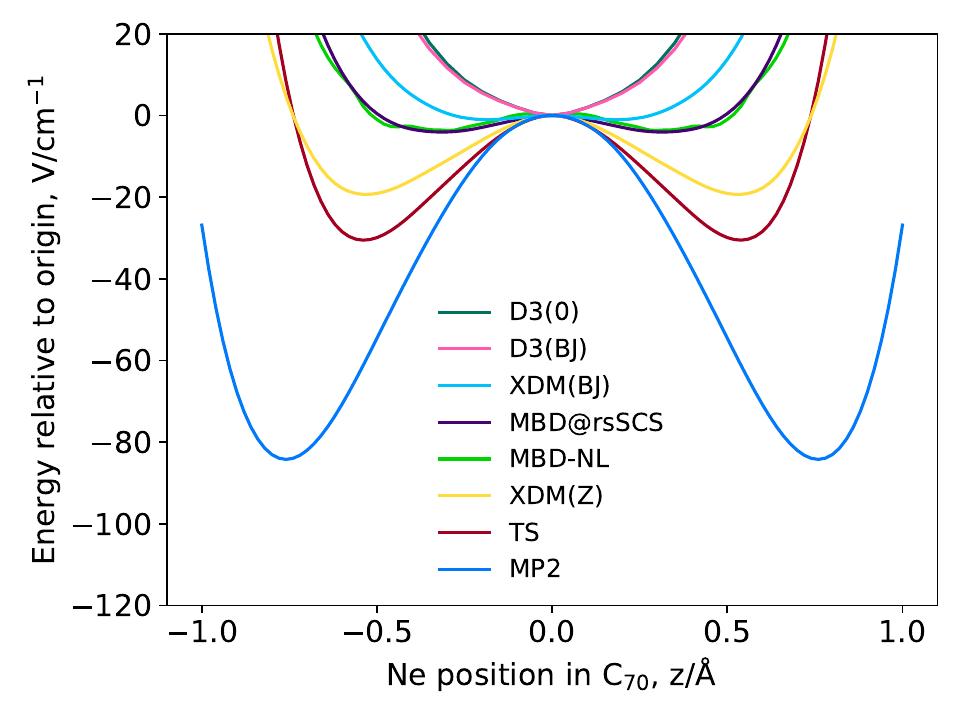} \label{fig: PBE0 dispersion}}
            \hfill
            \subfloat[DFA'+XDM]{\includegraphics[width=0.49\textwidth]{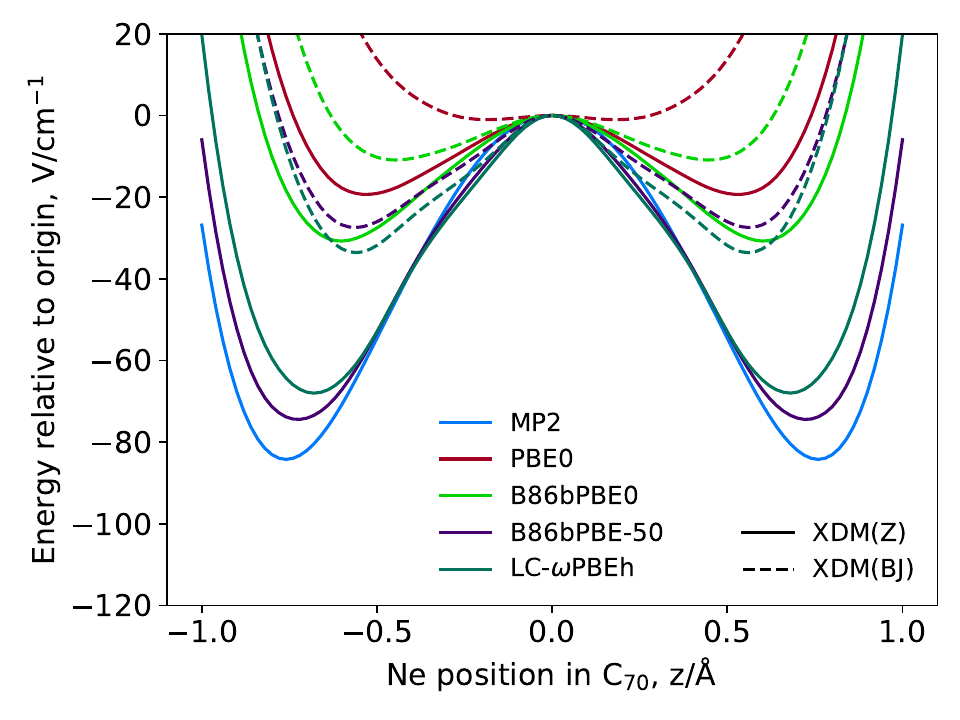} \label{fig: XDM PBE0}}\,
            \caption{PES slices of (a) PBE0 with dispersion corrections D3(0), D3(BJ), XDM(BJ), MBD@rsSCS, MBD-NL, XDM(Z), and TS in forest green, pink, turquoise, purple, green, yellow and maroon respectively; (b) XDM(BJ) as solid and XDM(Z) as dashed lines added on top of PBE0, B86bPBE0, B86bPBE-50 and LC-$\omega$PBEh base functionals in maroon, green, purple and forest green respectively. MP2 is shown on both panels in blue as a reference WF method for comparison.  
            }
            \label{fig: PBE0 Dispersion}
        \end{figure*}
        Next, we turn our attention to minimally empirical functionals designed for intermolecular interactions. In Figure \ref{fig: PBE0 Dispersion}(a), we consider PBE0 as the base functional to which an assortment of different DCs [XDM(BJ), XDM(Z), TS, MBD-NL, MBD@rsSCS, D3(0), and D3(BJ)] are added as described in Eq~\eqref{eq: DFT+disp}. PBE0 is chosen as corresponding damping parameters are readily available for each of these DCs. 
        Both D3-type corrections exhibit a single minimum, indicating that they are not capturing sufficient information about the strength of dispersion interaction to transform the PBE0 curve (which also has a single minimum) into a double well. Furthermore, the XDM(BJ), MBD-NL, and MBD@rsSCS curves display extremely flat double wells, with BHs of $<$5~\wavenumber\,, much smaller than what was previously observed from the correlated wavefunction data in Figure \ref{fig: WF}. XDM(Z) and TS  yield deeper double wells than seen with the other dispersion corrections, although the corresponding barrier heights are still much smaller in magnitude than the MP2 and DLPNO-WF results. It is important to realise that we are most interested in how the dispersion energy \textit{changes} as the PES is traversed,\cite{gouldAreDispersionCorrections2018} not just its absolute value, which will be relatively large as this is the major component of the binding of noble gas endofullerenes.
        

        While it may seem possible to choose any base functional, and subsequently pair it with any DC, these choices are not necessarily independent. There is an interplay between these quantities that is reflected by the sensitivity of the empirical damping parameters to the choice of base functional. These parameters are present in all post-SCF dispersion methods to damp the dispersion energy at short interatomic separations, and are fit for each choice of base functional using reference WF data for molecular complexes. This implies that, as the base functional captures a better description of non-bonded repulsion, the less the DC needs to be damped. Conversely, the dispersion term is more damped when paired with functionals that already mimic some dispersion-like binding through their exchange terms, or include a fraction of MP2 correlation as in double hybrids. Damping parameter optimization is further complicated as the reference data is not typically limited to only dispersion-bound systems and spans a broad range of intermolecular interaction types, including hydrogen bonding. This results in error cancellation between delocalisation error in the base functional\cite{bryentonDelocalizationErrorGreatest2023} and the DC in some cases, particularly for GGAs. 
        In general, an accurate base functional should be dispersionless and have minimal delocalisation error, to allow the DC to do the heavy lifting.\cite{priceRequirementsAccurateDispersioncorrected2021} 

        \begin{table}[t]
            \caption{Barrier heights and minima positions for different dispersion corrections applied to PBE0 calculations, as well as XDM results for other selected DFAs.}
            \label{table: PBE0 Dispersion}
            \centering
            \setlength{\tabcolsep}{8pt}
            \begin{tabular}{ccc}
                 \hline Method&BH/\wavenumber&$z_m$/\AA  \\
                 \hline 
                 PBE0-D3(0)              &      0 &    0 \\
                 PBE0-D3(BJ)             &      0 &    0 \\ 
                 PBE0-MBD-NL             & ~~3.67 & 0.30 \\
                 PBE0-MBD@rsSCS          & ~~4.07 & 0.32 \\
                 PBE0-XDM(BJ)            & ~~1.02 & 0.18 \\
                 PBE0-XDM(Z)             &  19.32 & 0.54      \\
                 PBE0-TS                 &  30.54 & 0.54 \\
                 \hline
                 B86bPBE0-XDM(BJ)        &  10.91 & 0.44 \\
                 B86bPBE50-XDM(BJ)       &  27.43 & 0.56 \\
                 LC-$\omega$PBEh-XDM(BJ) &  33.59  & 0.56      \\
                 B86bPBE0-XDM(Z)         &  30.75  & 0.60      \\
                 B86bPBE50-XDM(Z)        &  74.42  & 0.72     \\
                 LC-$\omega$PBEh-XDM(Z)  &  67.99  & 0.68     \\ \hline
                 RPA@PBE0                &  33.10 & 0.62 \\ 
                 MP2                     &  89.84 & 0.76 \\
                 DLPNO-WF  & 67--95 & 0.60--0.80 \\
                 \hline
            \end{tabular}\\
        \end{table}
        
        Table~\ref{table: PBE0 Dispersion} and Figure~\ref{fig: PBE0 Dispersion}(b) show results obtained with the XDM dispersion correction using two choices of damping function and four base functionals. In general, there is little difference ($<$10 cm$^{-1}$) between PBE0 and B86bPBE0 exchange. However, the BH increases by ca.\ 20-40 cm$^{-1}$ using Z damping relative to BJ damping. The BH also increases with increased exact-exchange mixing, which should improve treatment of non-bonded repulsion and reduce delocalisation error. Overall, only LC-$\omega$PBEh-XDM(Z) and B86bPBE-50-XDM(Z) provide reasonable agreement with the DLPNO-WF reference data.

    \section{Conclusion \label{sec: Conc}}

        In this paper, we have investigated the behaviour of a variety of functionals, as well as dispersion corrections at describing the symmetric double-well potential in the \Endo{Ne}{C70}  endofullerene along its unique, anisotropic direction. This system was chosen due to its challenging nature for a concordant description of its PES between a variety of both WF and DFA methods. Through our exploration of different DFAs, spanning through most rungs of Perdew's ladder, it is not the case that a higher rung functional necessarily performs better, and converging the numerical integration grid is crucial. The choice of dispersion correction and damping function is also an important one, with a range of barrier heights and minima positions predicted depending on the specific choice. This is also affected by the choice of base functional, due to the interplay of the damping term between the functional and dispersion correction. When considering hybrid functionals, the precise amount of HF exchange is also a key variable, due to both the composition of the data sets used to fit the functional-dependent dispersion damping parameters and the functionals' ability to describe non-bonded repulsion in \Endo{Ne}{C70}. 
        
         Overall, the \Endo{Ne}{C70} system is extremely challenging for current state-of-the-art ES methods. As shown, almost any BH and $z_m$ can be calculated, depending on the particular DFA and DC. From the DFT perspective, it may be a leap too far to expect accurate and concordant results when even the WF methods are not in agreement with themselves. However, we can conclude that 
         functionals developed with only thermochemistry in mind are not necessarily well suited for a good description of the PES. It may be more pertinent to use a dispersionless base functional that obeys the physics of the system (with as few empirically derived factors as possible) paired with an added dispersion correction, and perhaps also calculation of the correlation energy using virtual orbitals as in double hybrid functionals.

         In closing, we recommend \Endo{Ne}{C70} as a diagnostic system to test correlation methods or dispersion corrections when developing new cutting-edge DFT methods. We also call for more experimental data on this system, in order to further drive improvement of electronic structure methods.
    
    \section*{Supplementary Information}
        See the Supplementary Information for more precise computational details for all the DFAs and WF methods shown.

    \begin{acknowledgements}
        KRB and ERJ thank the Natural Sciences and Engineering Research Council (NSERC) of Canada for financial support and the Atlantic Computing Excellence Network (ACENET) for computational resources. ERJ also thanks the Royal Society for a Wolfson Visiting Fellowship.
        DG acknowledges funding by the Deutsche Forschungsgemeinschaft (DFG, German Research Foundation) -- 498448112. DG thanks J.~Kussmann (LMU Munich) for providing a development version of the FermiONs++ software package. DG further thanks H.~Laqua (UC Berkeley) for helpful discussions.
    \end{acknowledgements}


    \bibliography{letter}

\end{document}


\title[SI: \Endo{Ne}{C70}]{Supplementary Information: Application of Correlated-Wavefunction and Density-Functional Theories to Endofullerenes: A Cautionary Tale} 

    \author{K. Panchagnula*\orcidlink{0009-0004-3952-073X}}
    \email{ksp31@cantab.ac.uk}
        \affiliation{Yusuf Hamied Department of Chemistry, University of Cambridge, Cambridge, United Kingdom}
    \author{D. Graf \orcidlink{0000-0002-7640-4162}}
        \affiliation{Department of Chemistry, University of Munich (LMU), Munich, Germany}
    \author{K.R. Bryenton \orcidlink{0000-0001-5716-5314}}
        \affiliation{Department of Physics and Atmospheric Science, Dalhousie University, 6310 Coburg Road, Halifax, Nova Scotia, Canada}
    \author{D.P. Tew \orcidlink{0000-0002-3220-4177}}
        \affiliation{Physical and Theoretical Chemistry Laboratory, University of Oxford, South Parks Road, United Kingdom}
    \author{E.R. Johnson \orcidlink{0000-0002-5651-468X}}
        \affiliation{Yusuf Hamied Department of Chemistry, University of Cambridge, Cambridge, United Kingdom}
        \affiliation{Department of Physics and Atmospheric Science, Dalhousie University, 6310 Coburg Road, Halifax, Nova Scotia, Canada}
        \affiliation{Department of Chemistry, Dalhousie University, 6243 Alumni Crescent, Halifax, Nova Scotia, Canada}
    \author{A.J.W. Thom \orcidlink{0000-0002-2417-7869}}
        \affiliation{Yusuf Hamied Department of Chemistry, University of Cambridge, Cambridge, United Kingdom}

    \date{\today}
    \maketitle

    \section{Computational Details}
        \subsection{FermiONs++ Details}
            All the calculations using the PBE,\cite{perdewGeneralizedGradientApproximation1996, perdewGeneralizedGradientApproximation1997} B97M-V,\cite{mardirossianMappingGenomeMetageneralized2015} B97,\cite{beckeDensityfunctionalThermochemistrySystematic1997} $\omega$-B97,\cite{chaiSystematicOptimizationLongrange2008} $\omega$-B97X,\cite{chaiSystematicOptimizationLongrange2008} $\omega$-B97X-V,\cite{mardirossianOB97XV10parameterRangeseparated2014} $\omega$-B97M-V,\cite{mardirossianOB97MVCombinatoriallyOptimized2016} B2PLYP,\cite{grimmeSemiempiricalHybridDensity2006} and PWPB95\cite{goerigkEfficientAccurateDoubleHybridMetaGGA2011} functionals were carried out using the FermiONs++ program package\cite{kussmannPreselectiveScreeningMatrix2013,kussmannPreselectiveScreeningLinearScaling2015} developed in the Ochsenfeld group. Calculations employing functionals up to rung 4 were carried out using the def2-TZVPPD\cite{rappoportPropertyoptimizedGaussianBasis2010,weigendBalancedBasisSets2005} basis set together with the def2-TZVPP-RI-JK basis set for resolution of the identity Coulomb (RI-J) builds.\cite{kussmannHighlyEfficientResolutionofIdentity2021a} For rung 5 functionals the def2-QZVP\cite{weigendGaussianBasisSets2003} basis set together with the def2-TZVP-RI\cite{weigendRIMP2OptimizedAuxiliary1998,hattigOptimizationAuxiliaryBasis2005a} basis set for the RI-MP2 part was employed; the basis set for RI-J was unchanged. 
            All calculations were considered converged when the energy difference decreased below $1.0\times10^{-8}$ and the norm of the commutator $[\mathbf{F}, \mathbf{P}]$ was smaller than $1.0\times10^{-7}$.
            For the grid generation, we employed the partitioning scheme described in Ref.\citenum{laquaImprovedMolecularPartitioning2018a}. The radial grids were constructed using the M4 mapping described in Ref.\citenum{treutlerEfficientMolecularNumerical1995}. For the angular grids, we employed Lebedev-Laikov grids.\cite{lebedevQUADRATUREFORMULASPHERE1999}

            Each atomic grid is generally constructed as a product of radial ($\tau$) and angular ($\sigma$) grids, expressed as
            \begin{align}
                \mathbf{r}_{\text{atomic}} &= r_{\tau} \mathbf{r}_{\sigma} \\
                \omega_{\text{atomic}} &= \omega_{\tau} \omega_{\sigma}
            \end{align}
            Here, $r_{\tau}$ denotes the radius of the radial shell, $\mathbf{r}_{\sigma}$ specifies the position of the angular grid point on the unit sphere, $\omega_{\tau}$ denotes the radial grid weight, and $\omega_{\sigma}$ represents the angular grid weight. 
            Due to the increase in isotropy of the density moving towards the nuclei, the atomic grid is partitioned into three regions
            \begin{align}
                \tau_{\text{inner}} &\le \frac{n_{\text{rad}}}{3} \\
                \frac{n_{\text{rad}}}{3} &< \tau_{\text{middle}} \le \frac{n_{\text{rad}}}{2} \\
                \tau_{\text{outer}} &> \frac{n_{\text{rad}}}{2}
            \end{align}
            with the number of angular grid points increasing from inner to middle to outer. Furthermore, the number of radial points is increased for heavier elements. For more details and the exact specifications of the grids (denoted $g1$ to $g7$), the reader is referred to Ref.~\citenum{laquaImprovedMolecularPartitioning2018a}.
            To evaluate the VV10 correction,\cite{vydrovNonlocalVanWaals2010} the $g7$ grid was employed in all calculations. For the seminumerical evaluation of exact exchange,\cite{laquaEfficientLinearScalingSeminumerical2018,laquaHighlyEfficientLinearScaling2020a} we always employed the multi-grid denoted as $gm7$. 
            For the XC functionals shown in Fig.~1b of the main text, all used a (200 radial, 1454 angular) integration grid to ensure numerical stability. Convergence testing of the $\omega$-B97M-V PES slice with respect to both angular and radial grid sizes is shown in Fig.~\ref{fig: Grids}.

        \subsection{ORCA Computational Details}
            The calculations using the $\omega$-B97M(2)\cite{mardirossianSurvivalMostTransferable2018} functional were carried out with the ORCA software package.\cite{neeseORCAQuantumChemistry2020, neeseSoftwareUpdateORCA2022, kollmarPerturbationbasedSuperCIApproach2019, neeseSHARKIntegralGeneration2023, ugandiRecursiveFormulationOneelectron2023} The calculations employed the def2-QZVP basis set, together with the def2/J basis set for RI-J and def2-TZVPPD/C basis set for the RI-MP2 part. For the SCF, very tight settings were chosen (keyword VERYTIGHT). The grids were built using the DEFGRID3 keyword. The VV10 correction was evaluated self-consistently (keyword SCNL).

        \subsection{CCSD(T0) Computational Details}
            The CCSD(T0) calculations are paired natural orbital (PNO) calculations using PNO threshold $T=10^{-6}$ and $10^{-7}$, extrapolated to the complete PNO space (CPS) and complete basis set (CBS) limits from cc-pVDZ and cc-pVTZ calculations calculated using \textsc{TURBOMOLE}.\cite{franzkeTURBOMOLETodayTomorrow2023, balasubramaniTURBOMOLEModularProgram2020} Extrapolations used power laws of the form $T^{-\frac12}$ and $X^{-3}$. A single point triple-$\zeta$ calculation with PNO threshold of $T=10^{_7}$ took 3 days to run on 1 node, 48 cores requiring 200GB RAM and 2TB disk. Perturbative triples and the counterpoise correction are significant contributions, indicating tight PNO thresholds and large basis sets are required to achieve wavenumber accuracy.

            

        \subsection{FHI-aims Computational Details}
            
            The FHI-aims calculations were carried out using version 240920-1 of the code.\cite{blumInitioMolecularSimulations2009} The calculations made use of FHI-aims' hybrid DFT framework \cite{renResolutionofidentityApproachHartree2012, levchenkoHybridFunctionalsLarge2015, kokottEfficientAllelectronHybrid2024}, through its ELSI infrastructure \cite{yuELSIUnifiedSoftware2018}, real-space partitioning and parallelization \cite{havuEfficientONONIntegration2009}, and efficient resolution of identity (RI) approach \cite{ihrigAccurateLocalizedResolution2015} for linear scaling. The default \texttt{tight} basis settings were used, which are approximately equivalent in accuracy to the Gaussian cc-PVQZ basis set. Finer radial grids than the \texttt{tight} defaults were necessary in order to achieve a wavenumber-level of accuracy and eliminate PES oscillations from hybrids with high levels of Hartree-Fock exchange mixing. Convergence testing showed that a radial grid of 100 points was sufficient, set via keywords \texttt{radial\_base 50.0 7.0} combined with \texttt{radial\_multiplier 2.0}. 
            
            In the present work, dispersion corrections were paired with the PBE0 and B86bPBE0 hybrid functionals, both of which include 25\% exact (Hartree-Fock) exchange. 
            The PBE0 functional was selected due to its widespread use in the literature, while B86bPBE0 was selected as B86b is our preferred exchange functional for its ability to accurately describe non-bonded repulsion, as a result of the adherence of its enhancement factor to the appropriate large reduced-density-gradient limit. \cite{priceRequirementsAccurateDispersioncorrected2021} We also considered the B86bPBE50 hybrid functional (50\% exact exchange) and the LC-$\omega$hPBE range-separated hybrid (20\% short-range exact exchange and 100\% long-range exact exchange) with $\omega=0.2$.


            The XDM model calculates the dispersion energy via an asymptotic pairwise sum as
            \begin{equation}
                E_{\text{disp}}^{\text{XDM}} = - \sum_{i<j} \frac{f_{6} \, C_{6,ij}}{R_{ij}^{6}} + \frac{f_{8} \, C_{8,ij}}{R_{ij}^{8}} + \frac{f_{10} \, C_{10,ij}}{R_{ij}^{10}} \,,\label{eq: XDM}
            \end{equation}
            where $C_{n,ij}$ are the interatomic dispersion coefficients for atoms $i$ and $j$. $C_6$ captures instantaneous dipole-dipole interactions, $C_8$ captures dipole-quadrupole interactions, and $C_{10}$ captures both dipole-octupole and quadrupole-quadrupole interactions. To prevent unphysical divergence of the dispersion energy at short internuclear separations, the use of damping functions is required. XDM has traditionally used the Becke-Johnson (BJ) damping function:\cite{johnsonPostHartreeFockModelIntermolecular2005}
            \begin{equation}
                f_n^\text{BJ}(R_{ij}) = \frac{R^n_{ij}}{R^n_{ij} + R^n_{\text{vdW},ij}} \,,\label{eq: BJ damp}
            \end{equation}
            where 
            \begin{equation}
                R_{\text{vdW},ij} = a_1 R_{c,ij} + a_2 \,,\label{eq: vdW radius}
            \end{equation}
            is the van der Waals radius, defined in terms of $R_{c,ij}$ --- the ``critical'' separation where successive dispersion coefficients become equal. Due to overbinding of alkali-metal clusters obtained with XDM and BJ damping, Becke recently proposed a new damping function based on the atomic number, $Z$:\cite{dh24}
            \begin{equation}
            f_n^\text{Z}(R_{ij}) = \frac{R_{ij}^n}{R_{ij}^n + z_\text{damp} \frac{C_{6,ij}}{Z_i+Z_j}}
            \end{equation}
           We have recently implemented Z-damping in FHI-aims and have assessed the XDM(Z) method for a wide range of thermochemical benchmarks.\cite{xcdmz}
            The empirical parameters (either $a_1$ and $a_2$ or $z_\text{damp}$), are optimized by minimizing the root-mean-square-percent error (RMSPE) on the KB49 benchmark \cite{kannemannVanWaalsInteractions2010} of small molecular dimers for each basis set and functional combination.

    \section{Meta-GGA Oscillations}
        \begin{figure*}
            \centering
            \subfloat[(50, $a$) angular grids]{\includegraphics[scale=0.52]{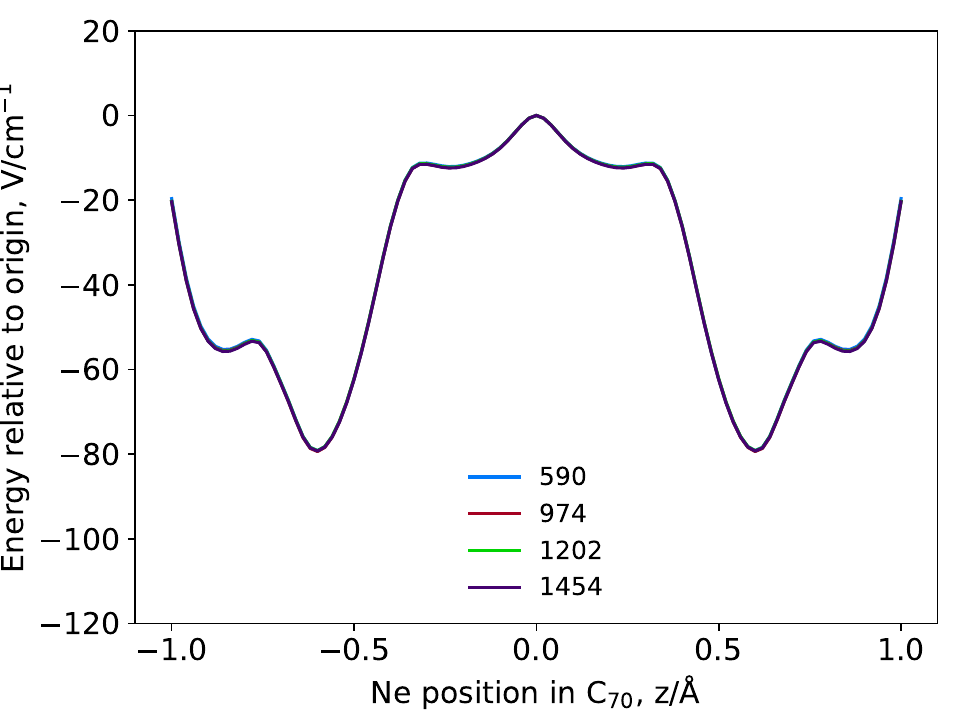}\label{fig: Angular Grid}}\,
            \hfill
            \subfloat[($r$, 590) radial grids]{\includegraphics[scale=0.52]{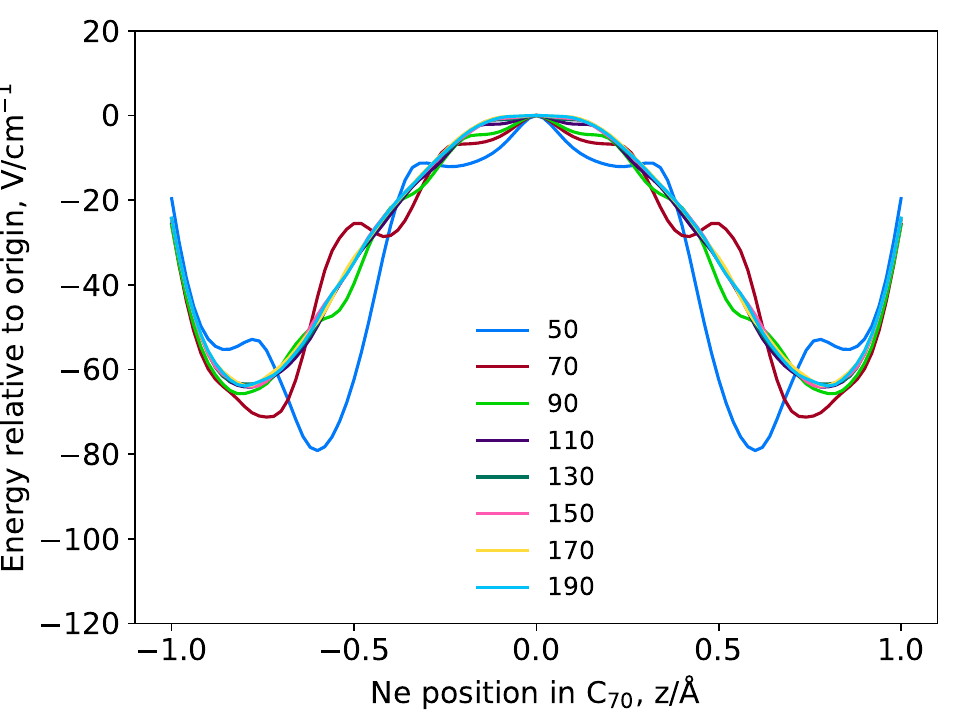}\label{fig: Radial Grid}}
            \caption{Effects of varying (a) angular and (b) radial grid sizes on the oscillatory behaviour of the PES for the $\omega$-B97M-V functional. Angular grids are (50, [590, 974, 1202, 1454]) in blue, maroon, light green and purple; Radial grids are ([50, 70, 90, 110, 130, 150, 170, 190], 590) in blue, maroon, light green, purple, forest green, pink, yellow and light blue.}
            \label{fig: Grids}
        \end{figure*}

        For the $\omega$-B97M-V functional, one-dimensional slices of the \Endo{Ne}{C70} PES are presented in Figure \ref{fig: Grids}, with Figure \ref{fig: Angular Grid} varying the angular grid size from 590 to 974, 1202, and 1454 grid points, each with 50 radial points. As the curves are almost completely overlapping, it is evident that the 590 Lebedev grid is sufficient for the numerical quadrature. However, the curve shape  demonstrates severe oscillatory behaviour, as has been commonly seen in the literature with mGGA functionals.\cite{johnsonPostHartreeFockModelIntermolecular2005, johnsonOscillationsMetageneralizedgradientApproximation2009} 
        Figure \ref{fig: Radial Grid} shows analogous results where the angular grid is fixed at 590 points and the radial grid size varied from 50 to 190 in increments of 20 points. As the quadrature is steadily increased, the region near the origin smoothens and flattens, with the minima around $\pm$0.75\AA\, revealing themselves. The smallest grid to give an acceptable PES is the (150, 590) mesh, but this may not be constant for all functionals.

    \bibliography{letter}